\documentclass[showpacs,aps,prb,reprint,groupedaddress,preprintnumber]{revtex4-1}

\usepackage{graphicx}
\usepackage{dcolumn}

\begin{document}


\title{Evolution of correlation strength in K$_{x}$Fe$_{2-y}$Se$_2$ superconductor doped with S}


\author{Kefeng Wang}
\author{Hechang Lei}
\author{C. Petrovic}
\affiliation{Condensed Matter Physics and Materials Science Department, Brookhaven National Laboratory, Upton, New York 11973 USA}

\date{\today}

\begin{abstract}
We report the evolution of thermal transport properties of iron-based superconductor K$_x$Fe$_{2-y}$Se$_2$ with sulfur substitution at Se sites. Sulfur doping suppresses the superconducting $T_c$ as well as the Seebeck coefficient. The Seebeck coefficient of all crystals in the low temperature range can be described very well by diffusive thermoelectric response model. The zero-temperature extrapolated value of Seebeck coefficient divided by temperature $S/T$ gradually decreases from $-0.48 \mu V/K^2$ to a very small value $\sim$ 0.03 $\mu$V/K$^2$ where $T_c$ is completely suppressed. The normal state electron Sommerfeld term ($\gamma_n$) of specific heat also decreases with the increase of sulfur content. The dcrease of $S/T$ and $\gamma_n$ reflects a suppression of the density of states at the Fermi energy, or a change in the Fermi surface that would induce the suppression of correlation strength.
\end{abstract}

\maketitle

\section{Introduction\label{Introduction}}
Superconductivity in pure and F-doped LaFeAsO with $T_c$ up to 26 K has opened a new frontier in the investigation of the novel superconducting materials and mechanisms.\cite{discovery,review1,review2} After an intensive study, superconductivity was discovered in several different type of iron-based materials, including REOFePn (RE=rare earth; Pn=P or As, 1111-type),\cite{1111-1,1111-2,1111-3,1111-4,1111-5} doped AFe$_2$As$_2$ (122-type, A=Ba, Sr, Ca),\cite{122-1,122-2,122-3} Fe$_2$As-type AFeAs (111-type, A=Li or Na),\cite{111-1,111-2} as well as anti-PbO-type Fe(Se,Te) (11-type).\cite{11-1,11-2} All have similar structure with the common FeAs-layer units. Most undoped compounds are stripe-like antiferromagnetic (spin density wave, SDW) metals, and the magnetic ordering comes about simultaneously with the structural phase transition from tetragonal to orhtorhombic unit cell. Experimental and theoretical studies suggests that the high-$T_c$ superconductivity in iron-based superconductors is influenced by proximity to SDW phase transition. The doping brings along charge carries that suppress the SDW ordering. It was suggested that the superconductivity may be established via inter-pocket scattering of electrons between the hole pockets and electron pockets, leading to the $s^{\pm}$ pairing.\cite{pairing1,pairing2,pairing3}

Thermal conductivity, thermoelectric power (Seebeck coefficient) and specific heat, give some useful information about the Fermi surface, electron-phonon coupling, pairing mechanism, and gap structure.\cite{TEP-1,TEP-2,TEP-3,TEP-4,thermalconduct2}. On the other hand, most iron-based superconductors exhibit significant Seebeck coefficient, similar to the  cobaltates (such as Na$_x$CoO$_2$) and cuprates which have been widely studied for potential application. \cite{application1,application2} Seebeck coefficient reaches about 90 $\mu V/K$ in RFeAsO$_{1-x}$F$_x$ and Co-doped FeSe, rather close to values observed in Na$_x$CoO$_2$.\cite{application3,application4}

Superconductivity with relatively high $T_c\sim 30$ K was recently reported in a new series of iron-based superconductors A$_x$Fe$_2$Se$_2$ (A=K,Rb,Cs,Tl). \cite{kfese1,kfese2,kfese3,kfese4,kfese5} These compounds are purely electron-doped and it is found that the superconductivity might be in proximity of a Mott insulator.\cite{kfese3} Only electron pockets were observed in angle-resolved photoemission experiment without a hole Fermi surface near the zone center in ARPES experiment. \cite{arpes1,arpes2,arpes3} Just like other iron-based superconductors,\cite{anionheight} superconductivity in A$_x$Fe$_2$Se$_2$ is sensitive to the Pn doping and anion height between Fe and Pn layers. Sulfur doping at Se sites suppresses the superconductivity \cite{kfeses} and induces a spin glass narrow bandgap semiconductor ground state for complete S substitution on Se sites. \cite{kfes} Here we report the evolution of thermal transport properties of iron-based superconductor K$_x$Fe$_{2-y}$Se$_2$ with sulfur substitution. Seebeck coefficient and the electron Sommerfeld term in specific heat, are suppressed. This implies the suppression of density of states at the Fermi energy and decrease in correlation strength. K$_x$Fe$_{2-y}$S$_2$ exhibits large Seebeck coefficient at high temperature range which is attributed to the thermally activated carriers over the narrow bandgap.

\section{Experimental\label{Experimental}}
Single crystals of K$_x$Fe$_{2-y}$Se$_{2-z}$S$_z$ were grown from nominal composition K:Fe:Se:S=0.8:2:2-$z$:$z$ with different S content, as described elsewhere. \cite{kfeses,kfes} The elemental analysis was performed using an energy-dispersive x-ray spectroscopy (EDX) in a JEOL JSM-6500 scanning electron microscope and following the actual sulfur contents are used. Electrical and thermal transport measurements were conducted in Quantum Design PPMS-9. The crystal was cleaved to a rectangular shape with dimension 5$\times$2 mm$^{2}$ in the \textit{ab}-plane and 0.3 mm thickness along the \textit{c}-axis. Thermoelectric power and thermal conductivity were measured using one-heater-two-thermometer setup which allowed us to determine all transport properties of the crystal with steady-state method in Quantum Design PPMS-9. The heat and electrical current were transported within the \textit{ab}-plane of the crystal oriented by Laue camera, with magnetic field along the \textit{c}-axis and perpendicular to the heat/electrical current. Silver paint contacts were made directly on the crystal surface providing both good thermal contact and low electrical contact resistance. Since air exposure exceeding 1 hour will result in the surface oxidization, the exposure to air of crystals was less than 20 minutes. The relative error in our measurement for both $\kappa$ and $S$ was below $5\%$ based on Ni standard measured under identical conditions.

\section{Results and Discussions}
Fig. 1(a) presents temperature dependence of thermal conductivity for K$_x$Fe$_{2-y}$Se$_{2-z}$S$_z$ in zero magnetic field from 2 K to 300 K. Thermal
conductivity for all crystals exhibits a peak between 15 K and 30 K (Fig. 1(b)). The peak position moves to higher temperature with the increase of sulfur doping (red arrows in Fig. 1(b)). It changes from $\sim$ 17 K in undoped crystal to $\sim$ 25 K in K$_x$Fe$_{2-y}$S$_2$, and is considered to entirely originate from phonon contribution.\cite{kefeng} This shift is due to the lattice contraction corresponding to smaller radius of sulfur ion.

\begin{figure}
\includegraphics [scale=0.65]{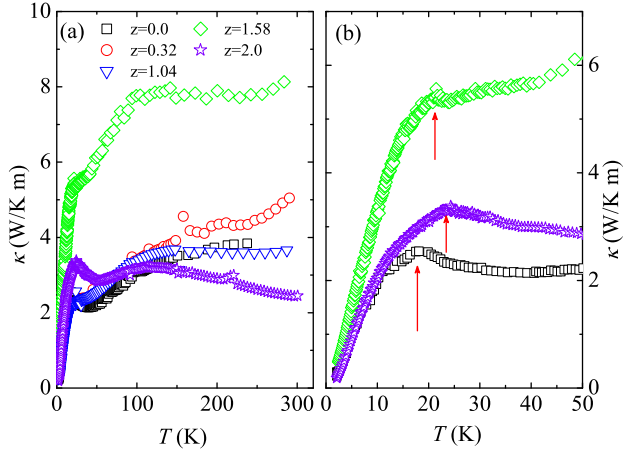}
\caption{(a) Temperature dependence of thermal conductivity for K$_x$Fe$_{2-y}$Se$_{2-z}$S$_z$ in zero magnetic field from 2~K to 300~K. (b) Low temperature thermal conductivity with phonon-related peak indicated by red arrows.}
\end{figure}

\begin{figure}
\includegraphics [scale=0.65]{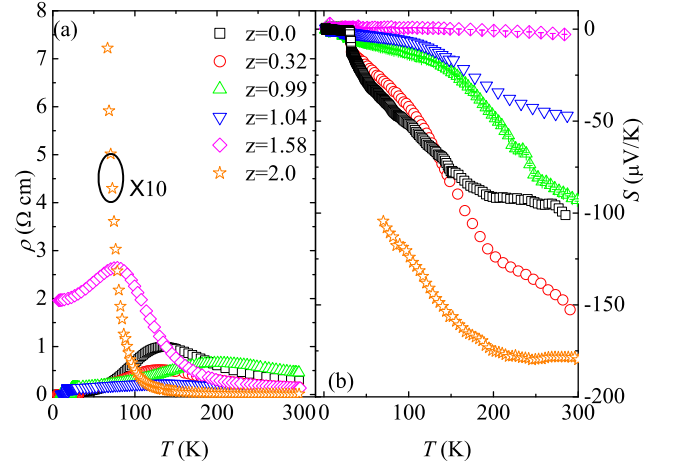}
\caption{Temperature dependence of resistivity (a) and Seebeck coefficient (b) for K$_x$Fe$_{2-y}$Se$_{2-z}$S$_z$ ($z=0.0,0.32,0.99,1.04$ and $1.58$) under zero magnetic field.}
\end{figure}


\begin{figure}
\includegraphics {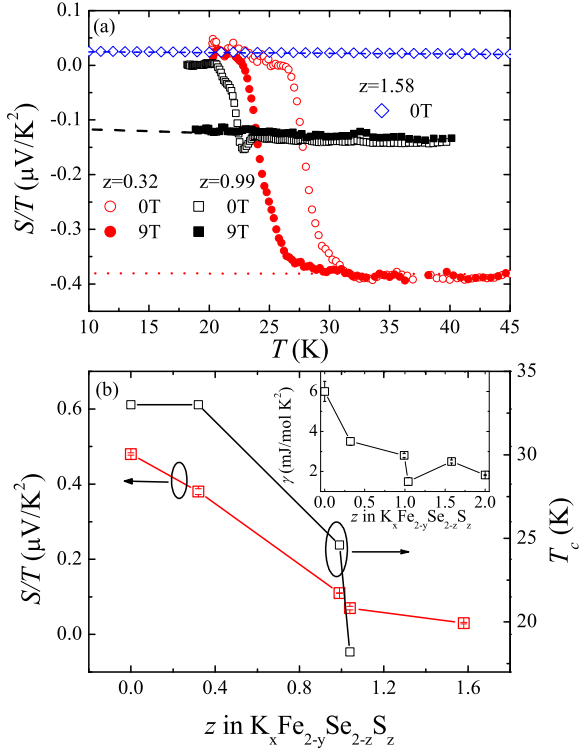}
\caption{(a) Temperature dependence of the Seebeck coefficient divided by \emph{T}, $\frac{S}{T}$, for K$_x$Fe$_{2-y}$Se$_{2-z}$S$_z$ with $z=0.32$, $0.99$ and $1.58$ under 0 T (open symbols) and 9 T (closed symbols), respectively. The dashed lines are the linear fitting results within high temperature range as described in text. (b) The relationship between the zero-temperature extrapolated value of $\frac{S}{T}$ (open circle) and superconducting $T_c$ (open square) to S concentration $z$ in  K$_x$Fe$_{2-y}$Se$_{2-z}$S$_z$. The inset shows the relationship between the Sommerfeld coefficient and S concentration $z$ in  K$_x$Fe$_{2-y}$Se$_{2-z}$S$_z$. For superconducting crystals, the Sommerfeld coefficients in normal state induced by magnetic field were used.}
\end{figure}

\begin{figure}
\includegraphics[scale=0.7] {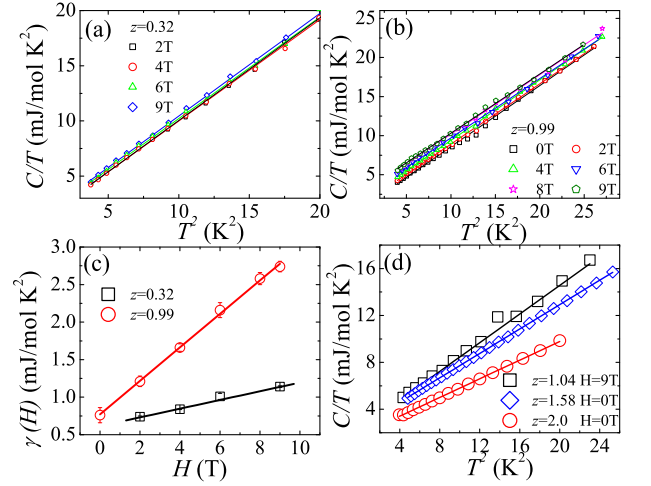}
\caption{(The specific heat data in low temperature range under different magnetic fields up to 9 T for crystals with $z=0.32$ (a) and $z=0.99$ (b). The lines are the linear fitting results. (c) Field dependence of the Sommerfeld coefficient $\gamma (H)$ for crystal with $z=0.32$ (squares) and $z=0.99$ (circles). (d)The specific heat data for crystal with $z=1.04$ under 9 T magnetic field, and for crystal with $z=1.58$ and 2.0 under 0 T field, respectively. The lines are linear fitting results.}
\end{figure}

With the increase of S, superconducting $T_c$ is suppressed and ultimately vanishes at $z=1.58$ (Fig. 2(a)). The suppression of $T_c$ is confirmed in Seebeck coefficient $S=0$ temperature (Fig. 2(b)) since Cooper pairs carry no entropy in the superconducting state. With increase in sulfur doping, the magnitude of Seebeck coefficient decreases significantly for superconducting samples. The Seebeck coefficient for crystal with $z=1.58$ (which does not exhibit superconducting transition above 1.9 K) is nearly zero in the whole temperature range (the value is $\sim$ $-0.75 \mu$V/K at 200 K and $\sim$ 0.06 $\mu$V/K at 2 K). For the narrow bandgap semiconductor K$_x$Fe$_{2-y}$S$_2$, $\rho$ and $S$ are beyond the detection limit of our instrument, as shown in Fig. 2(b). The observable Seebeck coefficient appears at $\sim$100 K and is due to the thermal excitation of carriers over the bandgap. With further increase of temperature, the number of the thermally excited carriers becomes larger and the Seebeck coefficient increases up to $\sim$180 $\mu$V/K$^2$ at 300 K.
For all crystals, there are no evident peaks in the Seebeck curves between 2 K and 300 K, indicating that there is no significant Fermi surface nesting for all crystals in this temperature range. \cite{TEP-1,TEP-3,kefeng}

\begin{table}[b]
\caption{Set of derived parameters for superconducting K$_{0.8}$Fe$_{2-y}$Se$_{-z}2$S$_z$ crystals.}
\begin{ruledtabular}
\begin{tabular}{ccccccc}
Parameter & z=0.0 & z=0.32 & z=0.99 & z=1.04 \\
$\frac{S}{T}$ ($\mu$V/K$^2$)& -0.48(3) & -0.38(8) & -0.11(5) & -0.13(7)  \\
$\gamma$ (mJ/mol K$^2$) & 6.0(5)$^a$ & 3.5(7) & 2.8(7) & 1.5(4) \\
$q$ & 0.13 & 0.09 & 0.12 & 0.15\\
$T_c$ (K) & 31.0 & 31.4 & 21.4 & 16.4  \\
$T_F$ (K) & 880 &1110 & 3860 & 3270 \\
$\frac{T_c}{T_F}$ & 0.04 & 0.028 & 0.005 & 0.005 \\
$m^*$ ($m_e$) & 3.4 & 2.7 & 1.1 & 1.3
\end{tabular}
\end{ruledtabular}
$^a$ The value is obtained from Ref.[44].
\end{table}


Seebeck coefficient in a material is the sum of three different contributions: the diffusion term $S_{diff}$, the spin-dependent scattering term and the phonon-drag term $S_{drag}$ due to electron-phonon coupling. \cite{TEP-1,text} TEP in our sample above $T_c$ is independent of magnetic field, which excludes the spin-dependent mechanism. The contribution of phonon-drag term often gives $\sim T^3$ dependence for $T<<\Theta_D$, $\sim 1/T$ for $T\geq\Theta_D$ (where $\Theta_D$ is the Debye Temperature), and a peak structure for $\sim\frac{\Theta_D}{5}$.\cite{text} The absence of the peak structure in our TEP results suggests negligible contribution of the phonon drag effect to $S(T)$ since $\Theta_D$ for crystals with z=0 and z=2.0 are 260 K and 289 K.\cite{kefeng,kfes}
At low temperature, diffusive Seebeck response of a Fermi liquid dominates and is expected to be linear in $T$ in the zero-temperature limit, with a magnitude proportional to the strength of electronic correlations.\cite{behnia} This is similar to the $T$-linear electronic specific heat, $C_e/T=\gamma$. In a one-band system both can be described by:
\begin{eqnarray}
S/T & = & \pm\frac{\pi^{2}}{2}\frac{k_B}{e}\frac{1}{T_F} =\pm\frac{\pi^{2}}{3}\frac{k_B^2}{e} \frac{N(\epsilon_F)}{n} \\
\gamma & = & \frac{\pi^2}{2}k_B\frac{n}{T_F}=\frac{\pi^2}{3}k_B^2N(\epsilon_F)
\end{eqnarray}
where $k_B$ is Boltzmann's constant, $e$ is the electron charge and $n$ is the carrier density, $T_F$ is the Fermi temperature which is related to the Fermi energy $\epsilon_F$ and the density of states $N(\epsilon_F)$ as $N(\epsilon_F)=\frac{3n}{2\epsilon_F}=\frac{3n}{k_BT_F}$.\cite{behnia} In a multiband system, this formula gives the upper limit of the Fermi temperature of the dominant band. The Seebeck coefficients of all crystals fit to this formula very well in low temperature range. Fig. 3(a) shows the relationship between the Seebeck coefficient divided by temperature ($S/T$) in K$_x$Fe$_{2-y}$Se$_{2-z}$S$_z$ with different S content under 0 T and 9 T magnetic field respectively. For superconducting crystals, the Seebeck coefficient in the normal state is independent of magnetic field and exhibits linear relationship with temperature in the low temperature range (Fig. 3(a)). The zero-temperature extrapolated values of $S/T$ for different crystals are shown in Fig. 3(b) and Table I. With sulfur doping, $\frac{S}{T}$ is suppressed from $-0.48 \mu V/K^2$ to a very small value $\sim$ 0.03 $\mu$V/K$^2$ for crystals without superconducting transition. Similar trend was observed in suppression of superconducting $T_c$ (Fig. 3(b)).

The specific heat measurements on all crystals can also give some useful information. Our crystals do not exhibit anomaly at superconducting transition, similar to previous report and possibly due to the very small superconducting contribution and the nodeless gap.\cite{hp} Yet the magnetic field dependent specific heat can yield important information about the Fermi surface. Fig. 4(a) and (b) shows the specific heat data plotted as $\frac{C}{T}$ vs $T^2$ in the low temperature region under various magnetic fields for crystal with $z=0.32$ and $z=0.99$ with upper critical field $H_{c2}
\sim 45$ T and $\sim 13$ T (for field applied along the $c$-axis of the crystals) respecively.\cite{hechang2} The magnetic field gradually enhances the specific heat, indicating the build up of the quasi-particle density of states. From linear fitting to $C/T$ vs $T^2$ (solid lines in Fig. 4(a) and (b)), we obtained linear dependence of Sommerfeld coefficient on the magnetic field (Fig. 4(c)). This is consistent with the results on K$_x$Fe$_{2-y}$Se$_2$ and the nodeless gap. \cite{hp} The slope of the line in Fig. 4(b)is $\sim$ 0.06(5) mJ/mol K$^2$T and $\sim $ 0.22(3) mJ/mol K$^2$ for two crystals with $z=0.32$ and 0.99 respectively. From them we estimate the value of normal-state electron specific heat coefficient $\gamma_n$ to be 3.5 mJ/mol K$^2$ using upper critical field $H_{c2}(0)\sim 45$T for $z=0.32$ and $\gamma_n=2.8(7)$ mJ/mol K$^2$ for crystal with $z=0.99$ using $H_{c2}(0) \sim$ 13 T respectively.\cite{hechang2} These values are smaller than the value ($\sim 6.0(5)$ mJ/mol K$^2$) in K$_x$Fe$_{2-y}$Se$_2$ system as shown in Table I and the inset of Fig. 3(b).\cite{hp}. Application of 9 T magnetic field parallel to $c$-axis completely suppresses the superconductivity in crystal with $z=1.04$ and we obtain $\gamma_n\sim 1.5$ mJ/mol K$^2$ directly from linear fit of the low temperature $C/T$ vs $T^2$ in 9 T (Fig. 4(d)). For the crystals with $z=1.58$ and 2.0, we derived the Sommerfeld coefficient from specific heat under zero field (Fig. 4(d)). The results are shown in Table I and Fig. 5(d). With increase of sulfur content the electron Sommerfeld coefficient in the normal state is gradually suppressed.

According to (1) and (2), $S/T$ and the electron Sommerfeld term in specific heat are related to the carrier density and the density of states at the Fermi energy. Since the sulfur has identical electronic configuration to selenium, there should be no change in the carrier concentration with sulfur doping because the elemental analysis is consistent with full occupancy of S(Se) sites.\cite{kfeses} The absolute value of the dimensionless ratio of Seebeck coefficient to specific heat $q=\frac{N_{Av}eS}{T\gamma_n}$ with $N_{Av}$ the Avogadro number, gives the carrier density. From the values of $S/T$ and $\gamma_n$ obtained previously, we derived $q$ values for four superconducting crystals (Table I). The $q$ values do not exhibit significant change. Therefore the suppress of $S/T$ and $\gamma_n$ reflects a suppression of density of states at the Fermi level or a change in the Fermi surface.

The ratio of the superconducting transition temperature $T_c$ to Fermi temperature $T_F$ gives information about the correlation strength in superconductors. The ratio $\frac{T_c}{T_F}\sim0.04$ for  K$_x$Fe$_{2-y}$Se$_2$ implies a weakly correlated superconductor.\cite{kefeng} With increase in sulfur content, the value of $\frac{T_c}{T_F}$ decreases as shown in Table I. This implies a suppression of correlation strength. The effective mass, $m^*$, derived from $k_BT_F=\frac{\hbar^2k_F^2}{2m^*}$, is also suppressed with increase of S content (Table I), consistent with  the decrease of correlation strength with S doping.


\section{Conclusion}
In summary, we studied the evolution of thermal transport and thermodynamic properties of K$_x$Fe$_{2-y}$Se$_{2-z}$S$_z$. The zero-temperature extrapolated value of Seebeck coefficient $S/T$  is gradually suppressed, and then undergoes a sharp decrease at $z=0.99$ to a very small value ($\sim$ 0.03(2) $\mu$V/K$^2$) for crystals with more sulfur content. The electron Sommerfeld term ($\gamma_n$) in specific heat also decreases with increasing of sulfur content. The suppression of $S/T$ and $\gamma_n$ reflects a suppression of density of states at the Fermi level or a change of Fermi surface near $\sim$ 40$\%$ sulfur doping level, possibly due to the change of the correlation strength. This coincides with the suppression of superconducting transition and shows that correlation effect might be important for superconducting $T_c$ in this material.

\begin{acknowledgments}
We than John Warren for help with SEM measurements. Work at Brookhaven is supported by the U.S. DOE under contract No. DE-AC02-98CH10886 and in part by the center for Emergent Supercondcutivity, and Energy Frontier Research Center funded by the U.S. DOE, office for Basic Energy Science.
\end{acknowledgments}


\end{document}